\documentclass[aps,pra,twocolumn,superscriptaddress]{revtex4}
\usepackage{amsmath}%
\usepackage{amsfonts}%
\usepackage{amssymb}%
\usepackage{graphicx}
\usepackage{comment}
\usepackage{mathtools,mathrsfs,color}
\usepackage{MnSymbol,bigints}
\usepackage{filecontents}
\usepackage{natbib}
\usepackage[titletoc, title]{appendix}
\usepackage[hidelinks]{hyperref}

\newcommand{\red}[1]{\textcolor{black}{#1}}

\newcommand{\dn}[1]{\textcolor{blue}{#1}}
\definecolor{grey}{gray}{0.5}

\newcommand{\degC}[1]{$^\circ$C}

\begin{document}
\title{Experimental Determination of Power Losses and Heat Generation in Solar Cells for Photovoltaic-Thermal Applications}


\author{Bruno Lorenzi}
\email[Corresponding author. E--mail: ]{bruno.lorenzi@unimib.it}
\affiliation{Department of Materials Science, University of Milano-Bicocca, via Cozzi 55, I-20125 Milan, Italy}
\affiliation{Mechanical Engineering Department, Massachusetts Institute of Technology, Cambridge, Massachusetts 02139, USA,}  
\author{Maurizio Acciarri}
\affiliation{Department of Materials Science, University of Milano-Bicocca, via Cozzi 55, I-20125 Milan, Italy}
\author{Dario Narducci}
\affiliation{Department of Materials Science, University of Milano-Bicocca, via Cozzi 55, I-20125 Milan, Italy}


\date{\today}

\begin{abstract}
Solar cell thermal recovery has recently attracted more and more attention as a viable solution to increase photovoltaic efficiency. However the convenience of the implementation of such a strategy is bound to the precise evaluation of the recoverable thermal power, and to a proper definition of the losses occurring within the solar device. In this work we establish a framework in which all solar cell losses are defined and described. Aim is to determine the components of the thermal fraction. We therefore describe an experimental method to precisely compute these components from the measurement of the external quantum efficiency, the current-voltage characteristics, and the reflectivity of the solar cell. Applying this method to three different types of devices (bulk, thin film, and multi-junction) we could exploit the relationships among losses for the main three generations of PV cells available nowadays. In addition, since the model is explicitly wavelength-dependent, we could show how thermal losses in all cells occur over the whole solar spectrum, and not only in the infrared region. 
This demonstrates that profitable thermal harvesting technologies should enable heat recovery over the whole solar spectral range. \\ \\
\textit{This is a pre-print of an article published in Journal of Materials Engineering and Performance. The final authenticated version is available online at:} \dn{\url{https://doi.org/10.1007/s11665-018-3604-3}}
\end{abstract}

\maketitle
\section{Introduction}
\label{introduction}

Photovoltaic (PV) technologies play a dominant role in  electric power generation using renewable resources, with PV market expansion and PV conversion efficiency improvements sustaining each other \cite{reportIEA}. Enhancements of the solar conversion efficiency are therefore highly desirable to promote further diffusion of solar converters \cite{polman2016}. 
A possible way to improve solar energy conversion comes from technologies combining PV devices with systems able to recover the heat unavoidably produced within solar cells. 
Co-generation of warm water or the use of thermoelectric generators (TEGs) provide typical examples \cite{yazawa2011,Tyagi2012,wang2011,park2013,hsueh2015,lorenziJMR}.
In all cases, the profitability of hybrid solar harvesters is limited by the requirement of keeping PV cells at the lowest possible temperature, as their efficiency decreases with temperature at a rate depending on the specific PV material. This is a very well-known hurdle in the making of effective hybrid solar cells, as reported in previous papers by the present authors \cite{narducci2016} and by other groups \cite{chow2010}. Reusing heat (to warm up air/water or to further convert it into electricity) may be then from completely counterproductive to quite profitable depending on the PV cell.

This paper aims at providing a practical, experimental tool to assess the convenience of hybridization in various types of PV cells.  
The method we present enables a detailed evaluation of the thermal power fraction (hereafter $\xi_{u}$) available in solar cells. 
With no need to refer to any specific use of the heat released by the PV cell, it will be shown that 
such a heat originates from the whole solar spectrum through the many mechanisms responsible for thermal losses occurring in the PV conversion process. This point is of utmost relevance, and may provide suitable guidance to strategies based on the solar--split approach and, more in general, to hybridization schemes using optical (radiative) coupling between the PV and the thermal stage of the harvester.

The experimental method just requires measurements of the external quantum efficiency (EQE), of the current--voltage (IV) characteristics, and of the reflectivity of the solar cell. Data are then elaborated in the framework of a model returning $\xi_{u}$ along with an evaluation of other (non-recoverable) losses. 

The method is validated on three types of solar cells, covering the current range of available PV technologies: a commercial silicon-based bulk solar cell, a lab-made thin-film solar cell made of Copper Indium Gallium Selenide (CIGS), and a commercial triple-junction solar cell (by Spectrolab).



\section{Theoretical framework}
\label{theory}
In a solar cell the unconverted fraction ($\phi_\mathrm{loss}$) of the incoming solar power can be defined as 
 	\begin{equation}
 	\label{lost}
	\phi_\mathrm{loss}=1-\eta_\mathrm{pv}=1-\frac{P_\mathrm{el}}{G A_\mathrm{abs}}
	\end{equation}
with $\eta_\mathrm{pv}$ the solar cell conversion ratio, $P_\mathrm{el}$ the output electrical power, $G$ the solar irradiance, and $A_\mathrm{abs}$ the cell area. 
The power loss  fraction is the sum of different kinds of losses. 
\red{We can sort them in four main classes:}
\begin{description}
\item[optical losses ($L_{1}$)], namely reflection losses ($L_\mathrm{1R}$), transmission losses ($L_\mathrm{1T}$), contact grid shadowing ($L_\mathrm{1sh}$), and absorptions which cannot generate charge carriers  ($L_\mathrm{1abs}$)
\item[source-absorber mismatch losses ($L_{2}$)] due to the under-gap portion of the solar spectrum ($L_\mathrm{2a}$), and carrier thermalization ($L_\mathrm{2b}$) accounting for the voltage drop to the conduction band edge
\red{\item[electron-hole recombination current losses ($L_{3}$)] which can be either radiative ($L_\mathrm{3rad}$) or non-radiative ($L_\mathrm{3Nrad-J}$), or due to electrical shunts $L_\mathrm{3sh}$}
\red{\item[electron-hole recombination voltage losses ($L_\mathrm{4}$)] which  accounts for the voltage loss associated to the $L_{3}$ class }
\end{description}
\red{Actually every $L_{3}$ loss has a voltage drop counterpart (cf.\ Appendix A for further details). 
These voltage drops are why solar cells exhibit voltages smaller than $E_\mathrm{g}/q$, and their sum actually accounts for the difference between $E_\mathrm{g}/q$ and voltage at maximum power $V_\mathrm{\red{mp}}$.} \\

\begin{figure}[t!]
		\includegraphics[width=1\columnwidth]{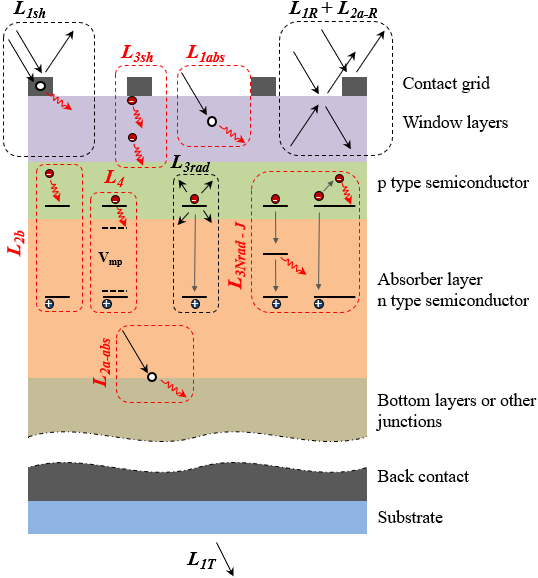}
		\centering
		\caption{Pictorial view of a general solar cell structure, with the losses occurring in it. Red squares marks thermal losses, black arrows the incoming radiation, and red arrows the losses. } 
		\label{fig1}
		\end{figure}

All the losses listed above contribute to set the cell conversion ratio: 
 \begin{equation}
	\eta_\mathrm{pv} = 1-\sum_{i=1}^4 L_{i} \equiv 1-\phi_\mathrm{loss}
	\end{equation}
A pictorial view of the loss mechanisms is reported in Fig. \ref{fig1}, where thermal losses are encircled in red. \\

Note that not all losses are converted into heat within the device. Therefore, the usable thermal fraction $\xi_{u}$ is smaller than $\phi_\mathrm{loss}$.
Specifically,  $L_\mathrm{1R}$ and $L_\mathrm{1T}$ are portions of the solar spectrum which are totally not absorbed, and thus do not contribute to $\xi_\mathrm{u}$.
In addition the contact grid can either absorb or reflect light, thus a portion of $L_\mathrm{1sh}$ can contribute to $\xi_\mathrm{u}$, while the remaining should be added to $L_\mathrm{1R}$. 
Considering the small contribution of the grid shadowing on the total device area, in this work we will make the assumption that all the light hitting the contacts will contributes to set the total reflection $L_\mathrm{1R-tot}$ (Eq. \ref{totalR}).

Regarding radiative recombination ($L_\mathrm{3rad}$) the photon generated by the recombination process either leaves the system or are re-absorbed, and eventually generate a electron-hole pair that is involved in a heat generation process. 
In this work we will consider all the photons generated by radiative recombination as emitted by the device and not re-absorbed. 
Thus $L_\mathrm{3rad}$ will contribute to the light reflected back by the device, setting $L_\mathrm{1R-tot}$ (Eq. \ref{totalR}).

Considering photon recycling negligible can be a source of error in evaluating thermal losses especially in the case of stacked multi-junction solar cells \cite{steiner2016, sheng2015, tex2016}. 
However in this work we will show that radiative recombination accounts for a very small fraction of the whole loss (1-3\%) showing how this assumption leads to marginal inaccuracies only.
In addition this approximation can be easily relaxed following Dupr\'e et al. \cite{dupre2016} considering a ratio for any of the recycling mechanisms that the emitted photon could encounter (leaving the cell, being absorbed by a process generating heat, or being absorbed by a process generating carriers). 
The problem with this approach is however to determine exact values for these ratios.

As of $L_\mathrm{2a}$, instead, since it cannot be absorbed by the absorber layer it is generally lost by \red{three mechanisms. It may be reflected (and thus contributes to $L_\mathrm{1R}$), or it is transmitted through the solar cell without interacting with it (and thus contributes to $L_\mathrm{1T}$), or it is absorbed by other cell layers (e.g.\ the window layers or the back contact) or by defects and traps, thus contributing to $L_\mathrm{1abs}$. 
Hereafter we will refer to these three mechanisms respectively as $L_\mathrm{2a-R}$, $L_\mathrm{2a-T}$ and $L_\mathrm{2a-abs}$. }
Thus, the total reflection and absorption losses can be written as
\begin{equation}
\label{totalR}
	L_\mathrm{1R-tot} = L_\mathrm{1R} + L_\mathrm{1sh} + L_\mathrm{3rad} + L_\mathrm{2a-R} 
	\end{equation}
while
	\begin{equation}
	\label{labstot}
	L_\mathrm{abs-tot} = L_\mathrm{1abs} + L_\mathrm{2a-abs}
	\end{equation}
\red{Thus the usable thermal power fraction reads
	\begin{equation}
	\label{thermal}
	\xi_\mathrm{u} = L_\mathrm{abs-tot}+L_\mathrm{2b}+L_\mathrm{3sh}+L_\mathrm{3Nrad-J}+L_{4}
	\end{equation}
or, alternatively, 
\begin{equation}
	\label{thermal_1}
	\xi_\mathrm{u} = 1-(\eta_\mathrm{pv}+L_\mathrm{1R-tot}+L_\mathrm{1T}+L_\mathrm{2a-T})
	\end{equation}}
In the following we will show how to quantify terms in Eq.\ \ref{thermal}, and the other losses as well, from the spectral analysis of the EQE, the reflectivity $R$, and the IV characteristics of the device.

\subsection{Quantum Efficiency}
\label{quantum_efficiency}
In the field of photovoltaics the EQE is defined as the ratio between the number of photons reaching the PV device and the number of electrons contributing to the output electrical current produced by the device. 
Experimentally, EQE can be obtained as
	\begin{equation}
	\label{EQE}
	EQE (\lambda) = \frac{I_\mathrm{out}(\lambda)}{I_\mathrm{ph}(\lambda)} 
	\end{equation}
where $I_\mathrm{out}(\lambda)$ is the device output current generated by a monochromatic radiation of wavelength $\lambda$, and $I_\mathrm{ph}(\lambda)$ is the current that the device would produce if all the incoming photons contributed to the device current.
Knowing the spectral dependency of the incident solar power, $I_\mathrm{ph}(\lambda)$ can be written as
	\begin{equation}
	\label{iph}
	I_\mathrm{ph}(\lambda) = \frac{q A_\mathrm{abs} G(\lambda)}{hc/\lambda} 
	\end{equation}
where $-q$ is the electron charge, $G(\lambda)$ is the spectral solar power density, 
$h$ is the Planck constant, and $c$ is the speed of light.


The internal quantum efficiency IQE($\lambda$) is instead the quantum efficiency without considering any optical loss, and can be written as 
	\begin{equation}
	\label{EQEIQE}
	IQE (\lambda) = \frac{EQE (\lambda)}{(1-R(\lambda))(1-T(\lambda))} 
	\end{equation}
where $R(\lambda)$ and $T(\lambda)$ are respectively the spectral device reflectivity and transmittance. 
In this work we consider only solar cells with opaque back contacts so that hereafter we will take $T(\lambda)=0$. 
However, the method may be easily extended to transparent back contacts (as often found in organic solar cells) by adding a measurement of $T(\lambda)$ to the characterization. 

Using Eqs.\ \ref{EQE} and Eq. \ref{EQEIQE} (with $T(\lambda)=0$) one immediately obtains 
\begin{equation}
	\label{igen1}
	IQE(\lambda)=\frac{I_\mathrm{out}(\lambda)}{I_\mathrm{ph}(\lambda)} \frac{1}{(1-R(\lambda))}=\frac{I_\mathrm{out}(\lambda)}{I_\mathrm{gen}(\lambda)}
	\end{equation}
where $I_\mathrm{gen}(\lambda)$ is the current that would be generated by the device if all photons actually entering the PV cell (thus those photons which are not reflected) will contribute to the device current. 

Using Eqs.\ \ref{iph}, and \ref{igen1} an explicit expression for $I_{gen}(\lambda)$ is obtained: 
  	\begin{equation}
	\label{igen}
	I_\mathrm{gen}(\lambda) = \frac{q G(\lambda)A_{abs}(1-R(\lambda))}{hc/\lambda} 
	\end{equation}
Finally, using Eq.\ \ref{igen} one can define the fraction of solar power actually entering the solar cell as 
	\begin{equation}
	\label{entering}
	G_\mathrm{gen}(\lambda) = G(\lambda)(1-R(\lambda))  
	\end{equation}

\subsection{Determination of Losses}
\label{losses_determination}
For the sake of clarity, it is useful to summarize the main assumptions made in the model.
\begin{enumerate}
\item the model neglects the photons that could be absorbed by the metallic contact grid and contribute to $\xi_{u}$, assuming that all photons hitting the contacts are reflected 
\item the model neglects photon recycling for radiative recombination, considering all these photons as emitted
\item the model takes into account only solar cells with opaque back contact, namely $T(\lambda)=0$\red{, and thus $L_\mathrm{1T}=L_\mathrm{2a-T}=0$}
\end{enumerate}
Losses may be now related to measurable quantities. \\
Since $R(\lambda)$ is defined as the whole device spectral reflectivity (thus accounting also for the contributions from $L_\mathrm{1sh}$, $L_\mathrm{2a-R}$, and $L_\mathrm{3rad}$) its relationship with the (integral) loss $L_\mathrm{1R-tot}$ is immediate, namely 
	\begin{multline}
	\label{LR}
	L_\mathrm{1R-tot} = L_\mathrm{1R} + L_\mathrm{1sh} + L_\mathrm{3rad} + L_\mathrm{2a-R} = \\  \frac{\int G(\lambda) R(\lambda) {\rm d} \lambda}{\int  G(\lambda) {\rm d} \lambda}
	\end{multline}
	
In addition the spectral dependency of $L_\mathrm{1R-tot}$ is simply given by 
	\begin{equation}
	\label{LR_lambda}
	L_\mathrm{1R-tot}(\lambda) = R (\lambda)
	\end{equation}
Likely conversions of spectral into integral quantities (and viceversa) may be carried out for all losses and  wavelength-dependent parameters.

Thus, using Eq.\ \ref{LR} for $L_\mathrm{1R-tot}$, and
Eq.\ \ref{entering} for $G_\mathrm{gen}(\lambda)$ one can actually calculate all remaining losses as follows. 

The under-gap fraction $L_\mathrm{2a}$ which contributes to $L_\mathrm{abs-tot}$ reads
 	\begin{equation}
	\label{el2a}
	L_\mathrm{2a-abs}(\lambda) = \frac{G_\mathrm{gen}(\lambda)}{G(\lambda)}H(\lambda - \lambda_\mathrm{g})
	\end{equation}
where $H(z)$ is the Heaviside step function
\begin{equation}
H(z)=\left\{
\begin{tabular}{ll}
1 &\mbox{for $z>0$}\\
0 &\mbox{otherwise}
\end{tabular}
\right.
\end{equation}
and $\lambda_\mathrm{g} = h c/E_\mathrm{g}$, with $E_\mathrm{g}$ the energy gap of the absorber material. 
This is clearly an approximation.
Actually, the absorbance of a semiconductor, especially for indirect energy gaps, is not a step function.
This leads to an underestimation of the thermal components coming from losses that involve the part of the solar spectrum with energy higher than the absorber material  $E_\mathrm{g}$ \red{(namely $L_\mathrm{2b}$, $L_\mathrm{3Nrad-J}$, $L_\mathrm{3sh}$, and $L_4$)}, and an overestimation of $L_\mathrm{2a}$, that depends upon the absorption of photons with energy lower than $E_\mathrm{g}$.

The carrier thermalization fraction $L_\mathrm{2b}$, accounting for the electron-hole relaxation to the band edge, is instead 
	\begin{equation}
	\label{el2b}
	L_\mathrm{2b}(\lambda) = \frac{G_\mathrm{gen}(\lambda) IQE(\lambda)}{G(\lambda)}  \left(\frac{\lambda_\mathrm{g}}{\lambda}-1\right)
	H( \lambda_\mathrm{g} -\lambda)
	\end{equation}

\red{A likely equation is valid for the sum of all the $L_\mathrm{4}$ losses} accounting for the relaxation between the band edge and the energy corresponding to the voltage at maximum power $V_\mathrm{mp}$, at which the solar cell is supposed to work:
\red{
	\begin{multline}
	\label{langle}
	L_\mathrm{4}(\lambda) = L_\mathrm{4carnot} + L_\mathrm{4boltz} + L_\mathrm{4Nrad-V} + L_\mathrm{4s} = \\ \frac{G_\mathrm{gen}(\lambda) IQE(\lambda)}{G(\lambda)} \left(1-\frac{qV_\mathrm{\red{mp}}}{E_\mathrm{g}}\right)	H( \lambda_\mathrm{g} -\lambda)
	\end{multline}
	}
\red{In Appendix A we show how to split the non-spectral contributions of every $L_\mathrm{4}$ component. 
}

\red{
The remaining losses can be only cumulatively estimated. 
Therefore we conveniently group them under the generic name of thermal losses $L_\mathrm{therm}$, computable as
	\begin{multline}
	\label{lrec}
	L_\mathrm{therm} (\lambda) =L_\mathrm{3Nrad-J} (\lambda) + L_\mathrm{1abs} (\lambda) + L_\mathrm{3sh} (\lambda) = \\ \frac{G_\mathrm{gen}(\lambda)  \left[ 1-IQE(\lambda)\right]}{G(\lambda)} 
	\end{multline}
	}
	

\red{Using Eq. \ \ref{labstot} and \ \ref{thermal}, along with Eqs.\ \ref{el2b}--\ref{lrec} one can determine the thermal fraction as a function of the wavelength 
(or in its integral form) by 
	\begin{equation}
	\label{thermal_losses}
	\xi_\mathrm{u}(\lambda) = L_\mathrm{2a-abs}(\lambda) + L_\mathrm{2b}(\lambda)+L_\mathrm{4}(\lambda)+L_\mathrm{therm}(\lambda)
	\end{equation}
}

A check of the impact of the approximations introduced in the model is achievable by computing $L_\mathrm{3rad}$.
\begin{figure}[t!]
		\includegraphics[width=1\columnwidth]{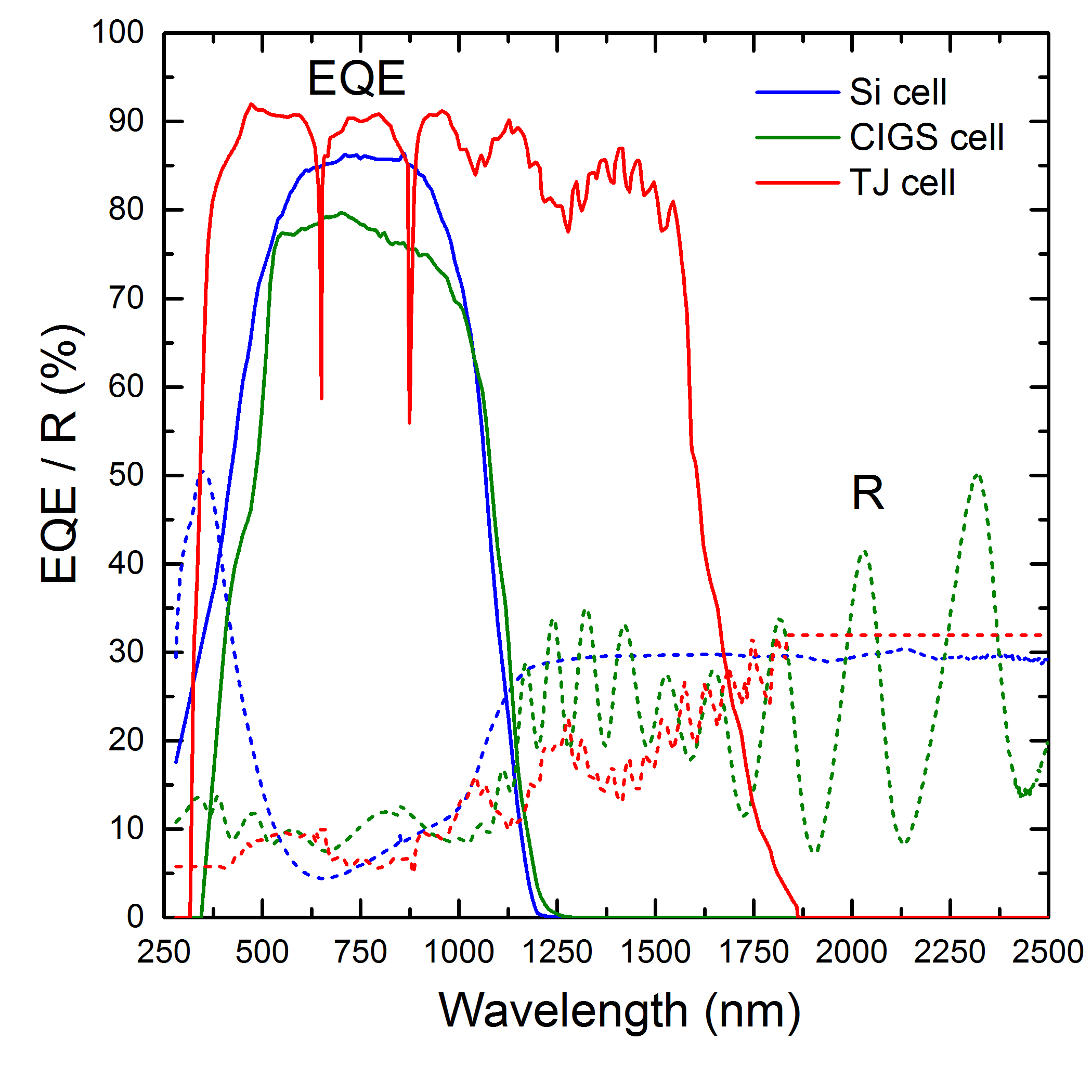}
		\centering
		\caption{$EQE(\lambda)$ and $R(\lambda)$ for the three solar cells analysed in this work. Data for the TJ solar cell were obtained from literature \cite{king2012}.} 
		\label{fig2}
		\end{figure}
Actually, considering that radiative recombination is basically the reverse of the optical absorption process, one may estimate the rate of the latter event, obtaining \cite{lorenziJMR}
	\begin{equation}
	\label{rad_rate}
	R_\mathrm{rad} = RR_{0} \left[\exp  \left( \frac{eV}{k_\mathrm{B}T}\right) -1\right] 
	\end{equation}
where $V$ is the external voltage, $k_\mathrm{B}$ the Boltzmann constant, $T$ the device temperature and  
	\begin{equation}
	\label{rr0}
	 RR_{0} =  \frac{2 \pi}{c^2 h^3} \int_{E_\mathrm{g}}^{\infty} \frac{E^2 {\rm d E}}{\exp \left[ E/k_\mathrm{B}T\right] -1}
	\end{equation}
\begin{figure*}[t!]
		\includegraphics[width=2\columnwidth]{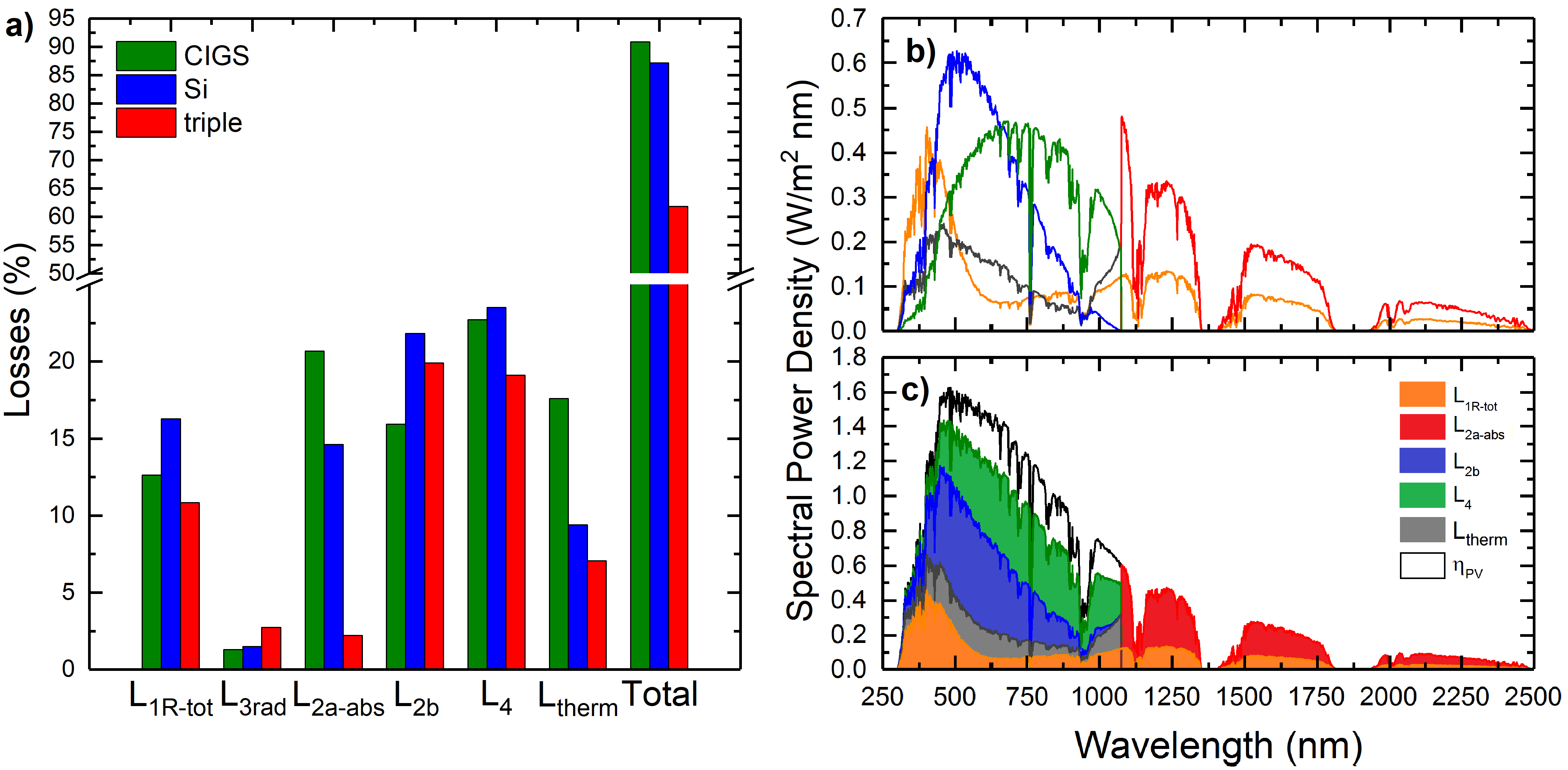}
		\centering
		\caption{(a) Computed losses  for the three solar cells. (b) Spectral dependency of losses in the Si cell. (c) Cumulative spectral dependency of losses for the Si cell compared to  the solar spectrum.} 
		\label{fig3}
\end{figure*}
In this work we will consider solar cells working at room temperature (300 K), but $R_\mathrm{rad}$ can be found at any temperature using Eq. \ref{rad_rate}.
The radiative recombination rate $R_\mathrm{rad}$ sets in turn the recombination current $I_\mathrm{rad}$. This leads to express $L_\mathrm{3rad}$ as
 	\begin{equation}
	\label{lrad}
	L_\mathrm{3rad} = \frac{I_\mathrm{rad}}{I_\mathrm{gen}} = \frac{q R_\mathrm{rad}A_\mathrm{abs}}{I_\mathrm{gen}}
	\end{equation}
that, in view of Eqs.\ \ref{igen} and \ref{entering}, becomes
	\begin{equation}
	\label{lrad2}
	L_\mathrm{3rad} (\lambda) = \frac{hc R_\mathrm{rad}}{\lambda G_\mathrm{gen}(\lambda)} 
	\end{equation}
		
\section{Materials and experimental}
\label{materials}
	In this work the losses of three different types of solar cells were evaluated. 
 	The first solar cell was a commercial, single-junction, bulk solar cell made of multicrystalline silicon (hereafter Si cell). 
 	The second solar cell was a lab-made single-junction thin film CIGS solar cell (hereafter CIGS cell). This cell was manufactured following a well-established procedure reported in a previous work \cite{acciarri2011}.
Both cells were measured using the same procedure and the same experimental setup. 
A SpeQuest Lot-Oriel quantum efficiency system was used to measure EQEs. Spectral response curves of PV devices were measured from 350 nm to 1800 nm with a 10 nm wavelength increment.
Current-voltage (IV) characteristics  were recorded under 1 Sun (100 mW/cm$^2$) illumination in Air Mass 1.5G conditions as generated by a Thermo Oriel Solar simulator.
Finally, $R(\lambda)$ was measured using a Jasco V-570 spectrometer equipped with an integrating sphere with a diameter of 60 mm between 250 and 2500 nm.

	The last solar cell was instead a commercial triple-junction GaInP\-/GaInAs\-/Ge solar cell (hereafter TJ cell) developed by Spectrolab, and the data needed for loss evaluation  were found in literature \cite{king2012}.
	
Figure \ref{fig2} reports  $EQE(\lambda)$ and $R(\lambda)$ data for Si, and CIGS cell, along with the data available for the TJ cell. 
Table \ref{Tab1} shows instead the efficiencies and the voltage at maximum power (obtained from I-V characteristics) along with the $E_{\rm g}$ values obtained from EQE measurements following a method reported in a previous publication \cite{marchionna2013}. 	

The procedure to access all loss terms is summarized for reader's convenience as follows:
\red{
\begin{enumerate}
\item inputting $R(\lambda)$ into Eq.\ \ref{entering} and making use of standard $G(\lambda)$ data one computes $G_\mathrm{gen}(\lambda)$
\item $L_\mathrm{1R-tot}$, is computed from Eq.\ \ref{LR}
\item $L_\mathrm{2a-abs}$ is then obtained from Eq.\ \ref{el2a}
\item $L_\mathrm{2b}$ follows from Eq.\ \ref{el2b}
\item $L_\mathrm{4}$ is computed from Eq.\ \ref{langle} 
\item $L_\mathrm{therm}$ is found from Eq.\ \ref{lrec}. 
\item the last contribution, namely $L_\mathrm{3rad}$ is calculated from Eq. \ \ref{lrad2}.
\end{enumerate}
}

\begin{table}[tb]
\centering
\begin{tabular}{rrrrrr}
		 			& CIGS	& 	Si		&	TJ 1	&	TJ 2	&	TJ 3	\\ \hline 
$\eta_\mathrm{pv}$ (\%)	& 10.03	&   12.89 	&	18.82	&	14.94	&	7.30	\\
$V_\mathrm{mp}$ (eV) 		& 0.39 	&   0.44  	&	1.31	&	1.07	&	0.40	\\
$E_\mathrm{g}$ (eV) 		& 1.25 	&   1.16  	&	1.89	&	1.41	&	0.67	\\
\end{tabular}
\caption{Values of $\eta_\mathrm{pv}$, $V_\mathrm{mp}$, and $E_\mathrm{g}$ for the three types of solar cells analysed in this work. }
\label{Tab1}
\end{table}

\section{Results and Discussion}
\label{results}
\begin{figure*}[t!]
		\includegraphics[width=2\columnwidth]{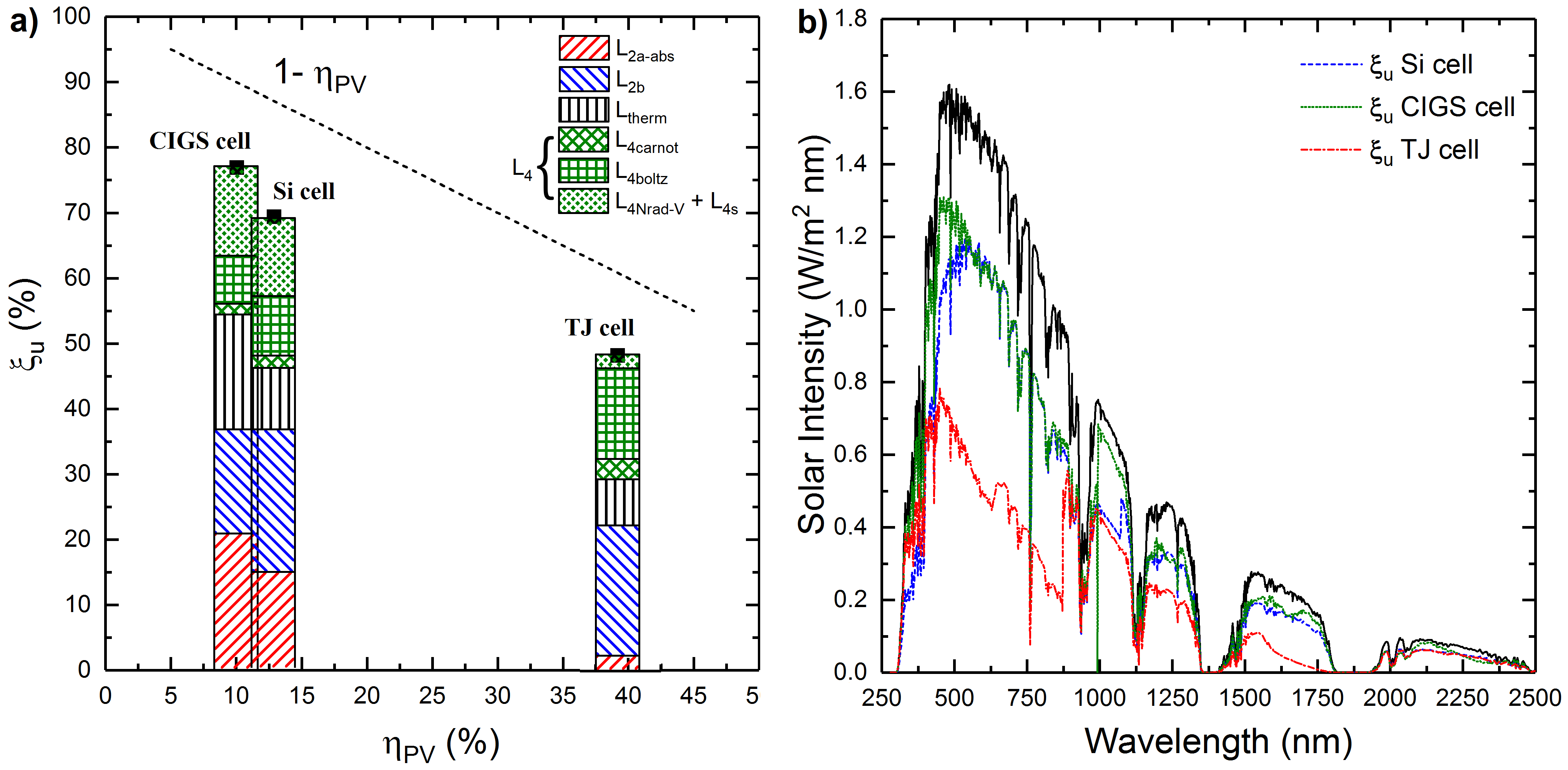}
		\centering
		\caption{(a) Bar graph of $\xi_\mathrm{u}$ and of its components vs.\ the cell efficiency for the three devices. The gap between the bars and the dashed line is the non-thermal lost power fraction. \red{See Appendix A for the $L_\mathrm{4}$ components.} (b) Spectral dependency of $\xi_\mathrm{u}$ for the three cells.} 
		\label{fig4}
		\end{figure*}
Figure \ref{fig3} a and Table \ref{Tab2} report the losses computed for the three solar cells.
\begin{table}[b!]
\centering
\begin{tabular}{c|c|c|c}
	 					&CIGS		&	Si		&	TJ	\\ \hline 
$L_\mathrm{1R-tot}$ (\%)		& 12.63		&   16.29 	&	10.83		\\
$L_\mathrm{3rad}$ (\%)			& 1.30 		&   1.48  	&	2.73	  \\
$L_\mathrm{2a-abs}$ (\%)		& 20.68 	&   14.63  	&	2.20	 \\
$L_\mathrm{2b}$ (\%)			& 15.94 	&   21.83  	&	19.90	  \\
$L_\mathrm{therm}$ (\%)	& 17.59 	&   9.39  	&	7.07	  \\
$L_\mathrm{4}$ (\%)		& 22.71 	&   23.52  	&	19.10	  \\
Total (\%)				& 90.86 	&   87.16 	&	61.80	\\
\end{tabular}
\caption{Values of computed losses.}
\label{Tab2}
\end{table}
As expected,  the total loss is higher for single-junction (CIGS and Si) solar cells. 
The sum of the total losses and of the cell efficiencies returns $\approx$100\% for all devices, with a maximum deviation of $\pm$ 1\%. 
This result validates the model and the suitability of the approximations it relies upon as well.

Figure \ref{fig3} a also clarifies that $L_\mathrm{2a-abs}$ and $L_\mathrm{therm}$ are mostly responsible for the loss differences among cells. 
Specifically, while the under-gap absorption loss $L_\mathrm{2a-abs}$ is almost negligible in TJ, in single-junction cells it is significant. 
This loss is found to be higher for CIGS because of its larger energy gap, and because of the presence of many layers on top of CIGS (buffer and finalization layers) \cite{acciarri2011} causing larger absorptions compared to the Si cell. 

Material quality rules instead $L_\mathrm{therm}$ which accounts for non-radiative recombination ($L_\mathrm{3Nrad}$), absorptions not generating carriers ($L_\mathrm{1abs}$), and electrical shunts ($L_\mathrm{3sh}$) -- all due to the presence of defects. 
Thus, the higher $L_\mathrm{therm}$ for CIGS is not surprising, and it actually witnesses the larger defectivity of the material.
Silicon and TJ solar cells are instead almost comparable, as the material quality is.

No relevant differences for the upper-gap losses, namely $L_\mathrm{2b}$ and $L_\mathrm{4}$ are found in the three types of solar cells \red{(we will highlight in Fig. \ref{fig4} a the differences about $L_\mathrm{4}$ components for the three solar cells analysed)}. 
For single-junction cells, CIGS shows the smallest losses, once again because of its higher $E_{\rm g}$. 
This is in line with what reported in previous works \cite{lorenziJMR}.
Interestingly enough, in the case of the TJ solar cell we found $L_\mathrm{2b}$ and $L_\mathrm{4}$ values very close to that of single-junction solar cells, as the addition of junctions cannot reduce these types of losses. This is consistent with previous evidence \cite{hirst2011} in the framework of the Shockley--Queisser limit \cite{SQlimit}.

Radiative loss $L_\mathrm{3rad}$ provides a marginal contribution, as expected. However it is interesting to note that it is larger for the TJ solar cell, as anticipated by Hirst et al.\ \cite{hirst2011} who correlated such an increase to the number of junctions.
The last contribution $L_\mathrm{1R-tot}$ mostly depends on the top layer roughness and on the anti-reflective coating used in the cell, so that it cannot be correlated to the absorber characteristics. \\
		
In summary, one may conclude that:
\begin{enumerate}
\item the material and device quality impact mainly on $L_\mathrm{therm}$;
\item For single-junction solar cells the energy gap set the balance between  $L_\mathrm{2a-abs}$ (that increases with $E_\mathrm{g}$) and $L_\mathrm{2b}$ (larger for smaller $E_\mathrm{g}$);
\item Multi-junction solar cells are very effective at limiting $L_\mathrm{2a-abs}$ but cannot avoid most of the $L_\mathrm{2b}$ and $L_\mathrm{4}$ contributions.
\end{enumerate}

Since all the losses were computed as a function of the wavelength, one may consider their  spectral dependence on the wavelength (Fig.\ \ref{fig3} b). 
The reported  case (Si cell) is representative of the trends observed also in the other cells.   
Figure \ref{fig3} b reports the spectral dependency of the losses calculated for the Si cell, while Fig.\ \ref{fig3} c shows their cumulative spectral dependency, with respect to the solar spectrum.

Concerning the thermal power loss, a plot of $\xi_\mathrm{u}$ vs.\ the cell efficiency $\eta_\mathrm{PV}$ (Fig.\ \ref{fig4} a) shows that  $\xi_{u}$ parallels $1-\eta_\mathrm{pv}$, rescaled by $\approx 10 - 15$\%. The downshift depends on $L_\mathrm{1R-tot}$ (cf.\  Eq.\ \ref{thermal_1}).
\red{Fig.\ \ref{fig4} a shows also the $L_\mathrm{4}$ components (see Appendix A). 
It is interesting to note how the total $L_\mathrm{4}$ loss, which is almost equal for all the cells, actually results from different combination of its components. 
In fact it can be seen how the higher radiative recombination in the TJ solar cell leads to a higher $L_\mathrm{4carnot}$, and $L_\mathrm{4boltz}$ contributions, which compensate the smaller ($L_\mathrm{4Nrad-V}$ +  $L_\mathrm{4s}$) component. 
For CIGS and Si solar cells instead $L_\mathrm{4}$ is basically equally split between ($L_\mathrm{4carnot}$ + $L_\mathrm{4boltz}$) and  ($L_\mathrm{4Nrad-V}$ +  $L_\mathrm{4s}$).}\\
From the spectral dependency of $\xi_\mathrm{u}$ showed in Fig.\ \ref{fig4} b, it is possible to see how the thermal fraction is quite equally distributed over the whole solar spectrum, and it is \emph{not} peaked in the infrared region.
Therefore, whichever  strategy is used to recover $\xi_\mathrm{u}$, it should be conceived so as to collect the widest spectral range.  
This leads to two rather important conclusions regarding spectrum splitting-based thermal recovery strategies, which are normally devoted to the harvesting of the infrared part of the solar spectrum \cite{vorobiev2006, kraemer2008, Mizoshiri2012}.
First, the use of such solutions in conjunction with multiple-junction cells may not be effective enough to justify the additional costs and complexity of the overall converter, as the harvester and the multi-junction policies compete to each other in the conversion of the long-wavelength part of the solar spectrum.
Second, they are necessarily sub-optimal, as the thermal power output is spread over the whole solar spectrum. Therefore, thermal harvesters should operate collecting heat at all wavelengths, covering also the short-wavelength region where heat resulting from carrier thermalization is larger.\\

It is worth stressing that these conclusions are limited to solar cells operating at room temperature. 
Clearly enough, at higher temperature the solar cell efficiency is expected to decrease \cite{virtuani2010} because of the increase of some losses. 
In particular, since the temperature sensitivity of solar cells is mainly due to a higher recombination ratio, $L_{\rm 3rad}$ and $L_{\rm 3Nrad-J}$ are expected to increase significantly, impacting consequently on $L_{\rm 4carnot}$, $L_{\rm 4boltz}$, and $L_{\rm 4Nrad-V}$ losses \cite{dupre2016}. 
A minor effect is instead expected on $L_{\rm 2b}$, and $L_{\rm 4s}$ losses, respectively due to the slight change of the energy gap of the absorber material, and to the associated small variation of the current flowing within the device.
		
\section{Conclusions}
\label{conclusion}
In this work we have reported a method to refine the evaluation of the usable thermal power released by solar cells. 
The method is based on a novel approach to the analysis of EQE, IV, and  reflectance measurements. 
It has been shown to be applicable to any kind of PV devices, and it is therefore very useful for a detailed evaluation of the thermal-recovery potential of a given solar cell.

Its application  to three different kinds of solar cells (bulk, thin film, and multi-junction cell) has shown that the material and device quality mostly set the thermal losses $L_\mathrm{therm}$. Also, it proved that in single-junction solar cells the energy gap modulates the balance between $L_\mathrm{2a-abs}$  and $L_\mathrm{2b}$. It was also shown that multi-junction cells are very effective at minimizing the $L_\mathrm{2a-abs}$ term, although they cannot significantly reduce $L_\mathrm{2b}$ and $L_\mathrm{4}$ losses.

Finally, the study  of the spectral  dependency of all terms  has shown how thermal losses are uniformly distributed over the whole solar spectrum, not only in the infrared region. This sets important constrains to viable thermal recovery strategies implementable when hybridizing PV systems.  

\section{Acknowledgements}
This is a pre-print of an article published in Journal of Materials Engineering and Performance. The final authenticated version is available online at: \dn{\url{https://doi.org/10.1007/s11665-018-3604-3}}. \\ \\

This project has received funding from the European Union's Horizon 2020 research and innovation programme under the Marie Sk\l{}odowska-Curie grant agreement No. 745304.

\begin{appendices}
\renewcommand{\thesection}{\Alph{section}}
\numberwithin{equation}{section}
\numberwithin{figure}{section}
\red{
\label{appendixA}
\section{Computation of $L_\mathrm{4}$ components}}
\red{In this section we show how to split the $L_\mathrm{4}$ components.
As mentioned in Sect.\ \ref{theory}, $L_\mathrm{4}$ losses are voltage drops associated to $L_\mathrm{3}$ losses.
Actually current losses $L_{3}$ impact the generation-recombination balance, reducing the voltage that the device can generate, and are the reason why solar cells exhibit voltages smaller than $E_\mathrm{g}/q$. 
The sum of these voltage losses actually accounts for the difference between $E_\mathrm{g}/q$ and voltage at maximum power $V_\mathrm{\red{mp}}$.\\
Previous studies \cite{henry1980, baldasaro2001, markvart2007, hirst2011, dupre2015} showed how the sum of two of such losses corresponds to the radiative recombination $L_\mathrm{3rad}$.
The first is called Carnot loss ($L_\mathrm{4carnot}$) with a corresponding voltage drop firstly derived by Landsberg and Badescu \cite{landsberg2000}:
 \begin{equation}
 \label{eq:DV_Carnot}
	\Delta V_\mathrm{4carnot} \approx \frac{E_\mathrm{g}}{q} \frac{T_\mathrm{c}}{T_\mathrm{s}}
 \end{equation}
with $T_\mathrm{c}$ the cell temperature, and $T_\mathrm{s}$ the temperature of the Sun. 
This loss takes into account only radiative emission in the solid angle within which the device absorbs the solar spectrum.  
The second is instead the so-called Boltzmann voltage loss ($L_\mathrm{4boltz}$) which takes into account the difference between the solid angle within which the solar cell absorbs the solar power, and the solid angle within which it emits.
The voltage drop associated can be calculated as
  \begin{equation}
  \label{eq:DV_Boltz}
	\Delta V_\mathrm{4boltz} \approx \frac{k_\mathrm{B} T_\mathrm{c}}{q} \mathrm{ln} \left(\frac{\Omega_\mathrm{emit}}{\Omega_\mathrm{abs}} \right) 
 \end{equation}
where $k_\mathrm{B}$ is the Boltzmann constant and $\Omega_\mathrm{emit}$ and $\Omega_\mathrm{abs}$ respectively the emission and absorption solid angles. \\
We then define with $L_\mathrm{4Nrad-V}$, and $L_\mathrm{4s}$ the voltage drops corresponding to non-radiative recombination, and to electrical shunts.} \\

\red{The total voltage drop due to $L_\mathrm{4}$ (hereafter $\Delta V_\mathrm{4}$) is therefore equal to
\begin{multline}
\label{L4_comp}
\Delta V_\mathrm{4} = \frac{E_\mathrm{g}}{q}-V_\mathrm{mp}= \\ \Delta V_\mathrm{4carnot}+\Delta V_\mathrm{4boltz}+\Delta V_\mathrm{4Nrad-V}+\Delta V_\mathrm{4s}
\end{multline}
While $V_\mathrm{mp}$ is known from the solar cell current-voltage characteristic, and the Carnot and Boltzmann contributions are known from Eqs.\ \ref{eq:DV_Carnot} and \ref{eq:DV_Boltz}, }
\red{one can obtain the sum of the two unknown voltage drops as
\begin{multline}
	\label{L4_nonrad}
	\Delta V_\mathrm{4Nrad-V} +  \Delta V_\mathrm{4s} = \\ \frac{E_\mathrm{g}}{q}-V_\mathrm{mp} - \Delta V_\mathrm{4carnot} - \Delta V_\mathrm{4boltz} 
	\end{multline}}
\red{Finally knowing from Eq. \ref{langle} the total $L_\mathrm{4}$, and from Eqs. \ref{eq:DV_Carnot}, \ref{eq:DV_Boltz}, and \ref{L4_nonrad}, the ratio between the different components, one can sort out the loss components $L_\mathrm{4carnot}$, $L_\mathrm{4boltz}$, and ($L_\mathrm{4Nrad-V} +  L_\mathrm{4s}$). \\
It is worth to point out that $\Delta V_\mathrm{4s}$ can also be extracted by the determination of the solar cell series resistance as 
 \begin{equation}
	\label{DV_4s}
	\Delta V_\mathrm{4s} = R_\mathrm{s} I_\mathrm{mp}
	\end{equation}
where $R_\mathrm{s}$ is the series resistance which can be obtained from the solar cell IV characteristic by several methods \cite{cotfas2012}, and $I_\mathrm{mp}$ the solar cell current at maximum power.}\\

Note that the method does not allow to obtain the spectral dependency of the $L_4$ components.

\end{appendices}

\newpage
\bibliographystyle{apsrev}
\bibliography{biblio}

\end{document}